# JOHN, THE SEMI-CONDUCTOR: A TOOL FOR COMPROVISATION

**Vincent Goudard**
Sorbonne Université , Collegium Musicæ
Paris, France
goudard@lam.jussieu.fr

**ABSTRACT**

This article presents "John", an open-source software designed to help collective free improvisation. It provides generated screen-scores running on distributed, reactive web-browsers. The musicians can then concurrently edit the scores in their own browser. John is used by ONE, a septet playing improvised electro-acoustic music with digital musical instruments (DMI). One of the original features of John is that its design takes care of leaving the musician's attention as free as possible.

Firstly, a quick review of the context of screen-based scores will help situate this research in the history of contemporary music notation. Then I will trace back how improvisation sessions led to John's particular "notational perspective". A brief description of the software will precede a discussion about the various aspects guiding its design.

## 1. INTRODUCTION

### 1.1 From traditional to graphical score

The score is generally considered as a tool for the composer to create a musical work for an interpreter. It describes the expected sonic result and prescribes the gestures to perform[1]. It thus stands as mnemonic mean to keep track of what is independent from the context of the performance[2] and often, is assimilated to the artwork itself in Western musical tradition.

Scores fulfill yet many other functions. It allows in particular to transpose the musical time into a visual space, enabling the composer to arrange musical elements "out of time" in order to produce pieces that could not be conceived without this visual support[3].

If the Western notational system invented by Guido d'Arezzo in the 11th century has continuously evolved, improving with new symbols and techniques until the early 20th century, the technological and cultural revolutions that followed subverted both the means of production and the range of musical expression, now extended to noise and the whole sound spectrum.

We can notice the development of so-called "graphic scores"[4] in the middle of the 20th century, that reflects this musical evolution for which the traditional notation is insufficient. For reasons that might seem opposite, the graphic score helped to push both the limits of what was possible to "fix" in a composition — by specifying it entirely on a synthesis system, and the limits of what it was conceivable to vary — the part entrusted to the performer's interpretation. The scores for Iannis Xenakis' *Mycene Alpha* and Earle Brown's *December 1952* highlight both these directions (see Figure 1).

This apparent opposition between a totally fixed work and a work that is totally subject to the performers' creativity seems more like the outcome of complementary approaches that aimed at exploring the new sound and musical domains, in their manifestations as well as in their potentialities, whether reified or fantasized.

Within this continuum of possibilities between fixed work and free improvisation, that Sandeep Bhagwati called "comprovisation" in [1], various notational perspectives can be considered. The various purposes of musical representation hitherto integrated in the traditional score gain independence and take a variable importance, adapting to the musical work and the performance contexts. The score defines the playing field, which is not necessarily linear and which, thanks to the possibility of producing animated images in real time, is no longer necessarily fixed.

### 1.2 Screen Scores

The increasing availability of digital devices led to the development of several applications meant for the creation of scores on-screen. As Lindsay Vickery notes in [20]:

> *These developments suggest a trend, particularly amongst young composers whose practice has developed exclusively on computer, to take the logical step to present notated materials on screen.*

---

[1] Eric Maestri proposes the terms "phonographic" and "ergographic" to describe both these aspects [14].
[2] Acknowledging here that the interpretation belongs to the contextual.
[3] A notorious example is the rondeau "*Ma fin est mon commencement*" (14th century) from Machaut in which the two voices are each other's retrograde.
[4] … that is, using graphic signs other than the usual symbols of the conventional notation of notes on a staff.

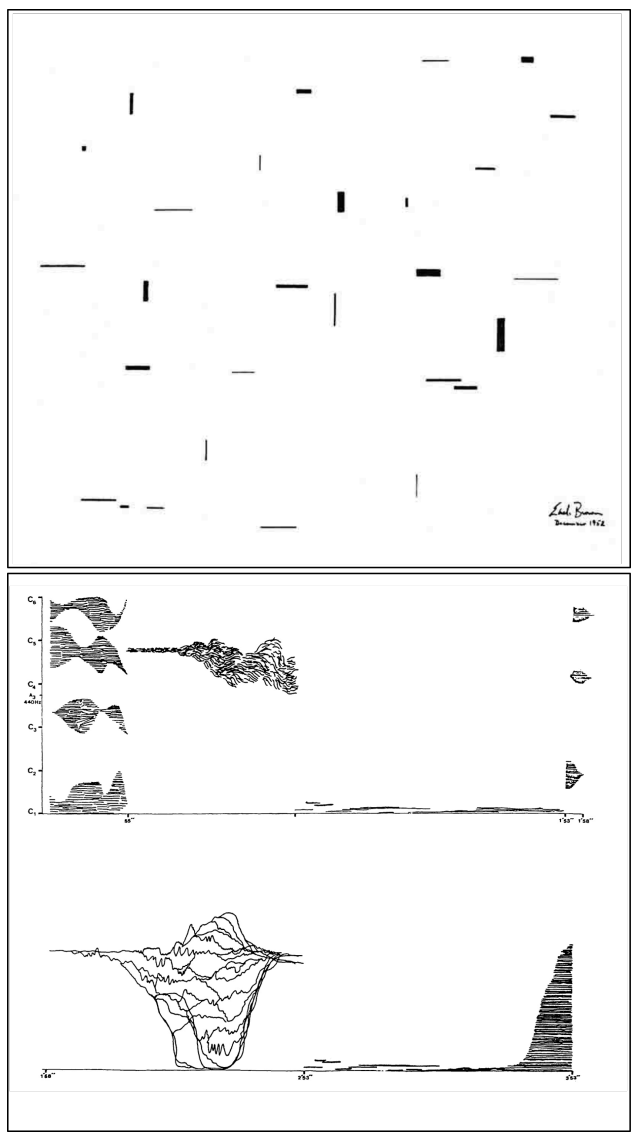

**Figure 1.** Extracts from *December 1952* by E. Brown (top) and *Mycène Alpha* by I. Xenakis (bottom).

Cat Hope summarizes the main features offered by this new medium with the following terms in [10]: *scrolling, permutative, transformative, generative and networking capabilities of the digital medium.*

Using computer graphics for the purposes of musical representation seems a medium of choice to enrich the possibilities of writing graphic scores. Especially, the fluidity of adaptation of the virtual medium makes it possible to envisage multiple "views" of the same score according to the contexts for which it is intended. Thus the composition, the performance, or the analysis of a musical work do not necessarily require the same representations. In terms of musical performance, we can add a distinction between the interpretation of a score by a human and a machine, these two types of "interpreters" affording relatively different abilities.

In the same way that digital technologies have atomized the musical instrument by decoupling its various constituents (gestural controller, mapping, synthesis, etc. becoming modular), they have also atomized the score into its various functions, supporting composition, performance or analysis. It is then necessary to specify which use case is at play and Cat Hope defines for this purpose the term "screen-score" as the medium presented to the musicians for a performance in [11]:

*Screen-scores are notated music compositions devised to be performed; and are not to be confused with visual representations of music or the musical interpretation of visual art.*

The concept of screen-scores has been investigated in depth by several authors, composers and musicologists, (in papers by Winkler [18], Clay [3] or Lee [12]) who discussed the advantages and drawbacks of using digital technologies for musical representation, both in its technical aspects and in its musicological consequences. Lindsay Vickery offers a very detailed review in [17] of critical latencies allowing an instrumentalist to read musical material displayed in real time and provides advices on what the composer should pay attention to when composing with this medium.

These studies offer relevant and valuable descriptions to the composer who wishes to work with screen-scores. However, it seems that they can be supplemented by a different approach to the score than those considered in most of the literature, in which the point of view is often that of the composer. The design of a screen-score system is consequently polarized by the central importance of the score, itself considered as a prerequisite for musical performance, a situation that also reflects a strong tradition of Western classical music[5].

In the case of ONE's performances (Figure 2), which are based on a practice of free improvisation devoid of prior composition, the focus moves towards the instrumentalist's side. The central element is not the score but the listening and understanding of sound and other musicians. The score (if it is still possible to call it so) often emerges after the improvisation sessions and its presence should not be at the expense of mutual attention. From this perspective, it is possible to envisage that the musician him/herself adapts the musical representation to his/her own needs, depending on the parts s/he has to play, her/his personal preferences, the various movements of the score, etc.

In the particular case where the instruments are digital and programmable, the use of a networked score system finally offers the possibility of delegating certain parameters of the instrument to an outsourced control supported by the score. In a situation of improvisation, the negotiation between this automated control and the choice of the musician implies a mediation that I will discuss later.

---

[5] A notable exception is the contribution of Georg Hajdu [9] who proposes the concept of "disposable music" to qualify musical forms "*that rely on a lesser degree on fully notated scores, such as "comprovisation" or laptop performance"*. However, even as "*disposable*" as it is, the score plays here again a prior role to the performance and remains central to the attention, differing from our approach.

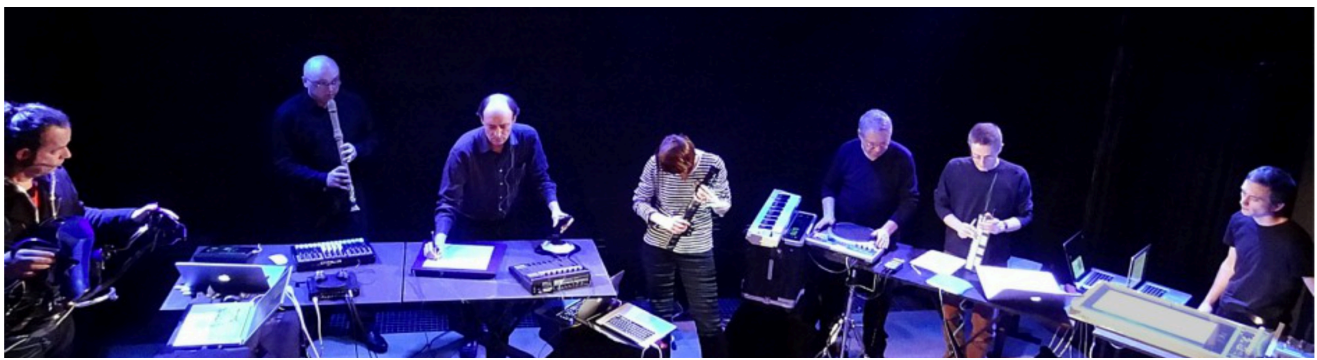

**Figure 2.** The members of ONE with their digital musical instruments.

### 1.3 Improvisation in ONE – birth of a notation

The seven musicians of ONE[6] are all deeply involved in the field of computer music with varying specialities applying in the fields of instrumental practice, composition, instrument making, research in music sciences and education. All of us practice digital musical instruments of which we built the software parts and sometimes also the hardware parts, to some extent. At the origin of our collaboration, there was no other project than that of attempting the experiment of playing a sound-based music together with such heteroclite digital instruments, without grid, without music theory, without prior agreement on the form and content.

Several improvisation sessions were opportunities to discover our sounds, our playing styles, our musical vocabulary. These moments of rehearsal were first and foremost the occasion of anarchic performances, guided only by the thread of our listening, to confront, to merge, to burst, to collide spaces, objects, soundtracks, along with moments of discussion and adjustments of our musical setup.

These sessions were also subject to classic improvisation exercises: searching for timbral fusion and counterpoints, fugal passages among musicians, accompanying a soloist, working on the pianissimo nuances, or playing "in the style" of a piece we knew. Eventually, audio recordings allowed us to play back the sometimes long and uninterrupted improvisations to extract interesting ideas.

The issue of large musical movements appeared before ONE's first public concert. The lack of a score structuring the concert's duration led us to follow a narrative scenario inspired by a novel by Jules Verne. Thus, the concert consisted of a series of chapters, simply identified by inter-titles in lieu of exotic and imaginary soundscapes to be explored.

Little by little, these experiences gave rise to the emergence of a more atomic musical vocabulary representing atmospheres and movements collectively defined, that we named "*karmas*"[7]. The various moments of play and discussion brought us to the development of other conceptual objects that were partly implemented in the form of a software nicknamed "John, the semi-conductor".

The origin of John's development is related to the desire to find a way to structure musical time in different movements within the perspective of freely improvised concerts of fairly long duration. Another motivation lies in the ability to vary the improvisations so as not to always repeat the same textures and formal structures such as sequences of ascending-descending cycles.

In addition, we were looking for ways to stimulate the exploration of unusual combinations and musical ideas pushing us out of our "comfort zone". The proposal to mathematically divide time into sequences to allow all possible ensembles of solo, duet, trio, up to the tutti, was the first impetus for the development of a score generator able to automatically produce such distributions.

As the opinions diverged within the band on the balance between rules and absence of rules, a key principle did find a ground of agreement: John is a semi-conductor. This means that scores created using John are just a proposition that each member of the group is free to follow or not, depending on the musical context that only takes shape during the very moment of the performance. Listening remains the primary rule of the game, prevailing over a blind follow-up of the score. In particular, the articulation between the different parts of the score, whether they are tiled or disjointed, or the act of playing when not supposed to, etc. is left to the appreciation of each musician.

This principle has the direct consequence of a streamlined design whose purpose is to allow each player to situate oneself within the score at a glance, without monopolizing her/his attention to the detriment of the other musicians. The goal is therefore very different from the one pursued in other screen-based musical notation systems such as those explored in works involving (extreme) sight-reading [8].

Essentially, John allows collective time management, whether during rehearsals, composition, or performances, providing a shared representation support. A brief description of the software to capture its outline will precede a discussion of the different aspects related to this group management.

---

[6] Performance except: https://youtu.be/lBVNwGeTxFA
[7] The relationship with this Indian concept is distant, but it does include an appealing meaning that echoes how we view them in performance: the set of actions represented by the karma influences the future of the individual. In the same way the musical interpretation of a *karma* (as we define it) is subject to the actions of the musicians and any accident, bifurcation with respect to the score will prevail on the musical evolution more than the score itself.

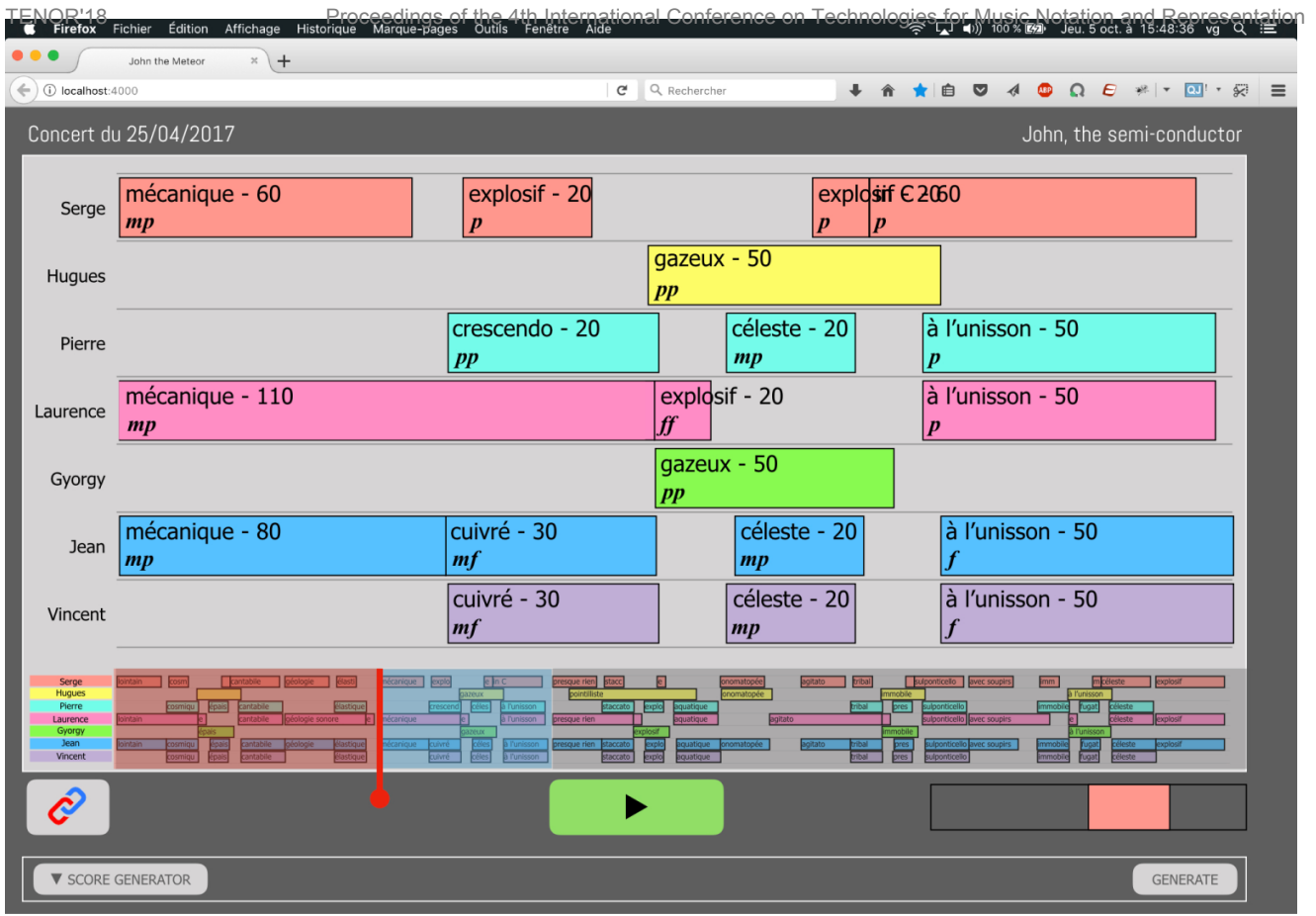

**Figure 3.** Snapshot of John's client interface.

## 2. ABOUT JOHN

John relies on a client / server architecture, in which each musician is visualizing a client interface in a web browser, on which s/he can act. This interface comes in two parts, a score generator on the one hand and an interactive visualization of the score on the other hand.

### 2.1 Score generator

The score generator makes it possible to very quickly create musical propositions by specifying only global constraints:

- overall duration of the score;
- minimum and maximum number of players;
- minimum and maximum duration of blocks;
- a list of *karmas* that identify a particular mood according to a common vocabulary established by the musicians during improvised practice sessions;
- nuances from pianississimo to fortississimo.

Once these constraints are specified, the score generator produces a random proposition that respects these conditions, that is composed of a sequence of time blocks associated with a *karma* and a nuance. This proposal can then be adjusted in the editing / viewing interface.

### 2.2 Interactive visualization

This interface represents blocks on a timeline. It consists of a reduced *global view* on one hand, giving a shared overview of the whole score, and a zoomed *local view*, located above the global view. On the global view is a playhead that is common and synchronous to all clients (in red on Figure 3), as well as a window (in blue on Figure 3) defining the time span displayed in the local view. This window is defined individually by each musician on their client and is typically ranging from ten seconds to a few minutes depending on the temporal granularity of the score and the preference of each.

All controls are accessible in all clients, so that anyone can edit the score: generating a new score, moving and changing the duration of the blocks and their content (*karma* and nuance), starting playback, changing the playback speed, moving the playhead to start at a given moment of the score. These changes will be immediately reported to all other John's clients.

The user can also define local parameters which will only affect her/his own client interface: the various tracks visibility, the duration of one's *local view* and the synchronization (or not) of one's local view to the reading cursor with the *link* toggle-button.

## 2.3 Implementation details

After a first version developed with Max[8], the application was brought to reactive HTML5 using the Meteor framework[9]. This allows collective editing on any platforms (including mobile platforms) connected to a LAN, via a simple web browser. The visualization was implemented using the D3.js library[10].

Scores are saved in JSON format as a list of events with a unique identifier, a track index, a start time, a duration, and a number of properties such as karma and nuance. During playback, time and score events are sent as MP messages [9] over the network using OSC[11].

## 3. JOHN IN PRACTICE

### 3.1 Generative composition

The score generator saved a lot of time during the rehearsals, giving us an immediate possible musical structure like an empty shell. As arbitrary as it is, its main function is to stimulate the musical performance with the most minimal prescription: play (or don't). Thus the proposals are often tried as they come before being adjusted collectively according to what members of the band find interesting or not. We can then evolve this musical structure, with apparently more efficiency than if we were to start out from nothing.

### 3.2 Distributing participation

The fact that John explicitly proposes a distribution of playing time among each musician has led to situations of performance that we would not necessarily have tried, especially in reduced configurations (solo and duet), each of us having a tendency to play too readily to actually leave room for these minimal configurations to settle.

Moreover, having "out of the game" moments makes it possible to better anticipate one's appearances. Indeed, digital musical instruments often have a "meta" dimension[12] and more generally a huge number of settings. These planned moments of rest make it possible to better manage the time we have to reach less accessible settings.

### 3.3 Tight synchronisation

At a micro-temporal level, synchronisation is impeded by the lack of idiomatic rules[13]. In particular, the absence of pulse or metric system makes the synchronization among musicians ever more difficult as their number increases and often deprives freely improvised musical forms from clear and tight transitions in large ensembles.

The conductor, when there is one, provides accurate cues, beats, and potential directions for play. Beside the ethical issues raised by the role of a leader in an improvisation band, raised by Canonne in [2], entrusting the conducting to a person[14] remains limited by the fact that s/he can only act in the present, and that it requires the almost permanent attention of the musicians, to the detriment of the attention they could bring to their peers. In this respect, the representation offered by John condenses in a certain way the score and the conductor in a single visual medium. The animated score ("scrolling score" in our case) offers indeed visual cues that indicate the simultaneity of several musical events, and its scrolling under the playhead allows a precise synchronization among the musicians at transitions.

### 3.4 Visual support for musical landmarks

Despite the availability of analysis tools[15] and a certain lexicon to describe sound and musical objects in electroacoustic music[16], there is no standard of prescriptive notation for digital musical instruments. The lack of a unanimous vocabulary, the singular nature of the instruments and the tremendous sound palette they provide can make it a nightmare to identify and discuss what has just been played during a long improvisation session (somehow failing here to use the word "rehearsal"). A minimal score such as that proposed by John facilitates this identification and allows to re-work specific moments after a long performance. The reduction that symbolic notation carries out on the complex sonic outcome of a performance allows everyone to quickly find one's way in the temporal space of an improvisation, faster than it would if we had to refer to the sound recording.

### 3.5 An ecology of attention

Free electroacoustic improvisation involves strong musicians' attention to other musicians, to their instrument and, obviously, to sound. In this respect, digital instruments often present the additional drawback, as compared to acoustic instruments, of capturing some of the visual attention due to the frequent presence of a screen, many interaction parameters, and an interface sometimes lacking tactile feedback or touch marks that would allow to access them without needing to look at them. Furthermore, digital musicians will often prepare their instrumentarium just prior to the performance with a chosen set of ad-hoc musical elements[17] (when not re-coding the whole thing) which further complicates a perfect knowledge of the ergonomics of the instrument, which would do without the visual.

John's design has been driven by an economy of cognitive load for musicians. Being able to partly customize one's visualization interface thus does not mean to add more visual data to it, but to see only what is necessary for one to gain collective awareness.

---

[8] https://cycling74.com/
[9] http://meteor.com/
[10] https://d3js.org/
[11] Open Sound Control: http://opensoundcontrol.org
[12] That is, it can be totally reconfigured during the performance to offer a whole other set of sounds, processes and playing modes.
[13] such as chord grids, time signature, scales, etc.

[14] such as using Walter Thomson's Soundpainting or in a composition like John Zorn's *Cobra*.
[15] Such as E-Analysis [5] or the GRM Acousmographe [7].
[16] In the work of Schaeffer [18], Bayle [2] or MIM [17] among others.
[17] In an informal discussion, Thor Magnusson used the term "pre-gramming" for this particular work that precedes a concert.

### 3.6 A score for humans and for (digital) instruments

During the score playback, the server sends data to clients as events begin or end (Figure 4). This information can be used by the musician's instrument (according this DMI is connected to the network). But, as John is only a "semi-conductor", it may as well be subject to the musician/client approval to allow some flexibility in the way the musician sticks to the score. Thus, it could have been devised that a specific *karma* recalls a corresponding preset of sounds in the musician's instrument, suitable for the *karma's* mood. But if the musician is still playing a previous other *karma*, s/he probably will not want this notification to automatically change the preset before s/he finishes one's current phrase. This "loose control" makes John's usage a little different from traditional sequencers.

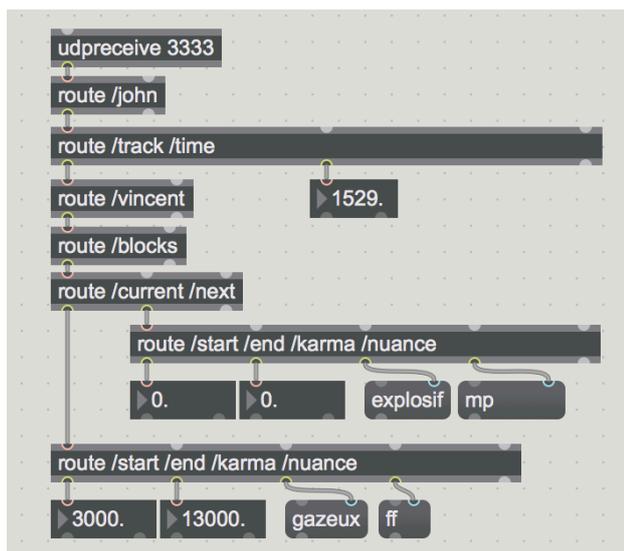

**Figure 4.** Receiving data in a Max instrument.

### 3.7 Showing the score?

Being able to read the interactions among the musicians in improvisation performances can contribute to the performance overall appreciation. Yet with DMIs, the spatial and energy decoupling between the instrumentalist's gestures and the resulting sound energy and location (on a possibly remote loudspeaker) confuses this reading. Screen-scores systems allow to share the score display with the audience more easily than printed scores and could help in this situation with the risk however, that it may "*detract from the dramatic performative aspects of the work*" among other reasons suggested by Cat Hope in [11]. Although John's score was never shown directly to the audience for this very reason, it has been used to control visual effects and lightings, meant both for stage design purpose and for helping listening and understanding of the music[18].

### 4. PERSPECTIVES

It has been acknowledged by the members of ONE that John was helping our creative process. However, there remains open issues like collective synchronization over rhythmic passages. Especially, anticipating a dynamic process is no trivial task and would probably require specific tools for that purpose, such as the animations proposed by Ryan Ross Smith in [19].

The concept of *local* and *global view* could probably be generalized to other shareable parameters. For example, being able to start a local playback in order to practice or prepare one's instrument on one's own. Similarly, it would be useful to work on another score than the ones loaded on others' clients. This de-synchronization raises however issues of versioning conflicts.

John's porting to a web technology is partly motivated by the possibility of future concerts involving a large number of musicians and where ever musician would be able to see his part with a simple web-browser. More developments will be needed to achieve such a performance, but there should not be technological locks.

Overall, computer-based scores give way to many possible interactions during performance time. Maybe the score should be considered as a collective instrument, which every musician and possibly the audience too, could play.

**Acknowledgments**

This work was partly carried out within a doctoral program supported by the Collegium Musicæ[19]. ONE is produced by Puce Muse[20]. Also, I would like to thank my colleagues from ONE for their valuable collaboration in the development of John : Laurence Bouckaert, Pierre Couprie, Hugues Genevois, Jean Haury, György Kurtág Jr. and Serge De Laubier.

---

[18] Examples include switching spotlights on musicians supposed to play, changing light hue according to the karmas, projecting aggregated sound waves as traces of the score, synching video, etc.

[19] http://collegium.musicae.sorbonne-universites.fr/
[20] http://pucemuse.com